\documentclass[12pt]{article}
\usepackage{eurosym}
\usepackage{amssymb,amsmath,epsfig}
\usepackage{color}
\usepackage{amsmath,amsfonts,amssymb}
\usepackage{geometry}
\begin{document}
\title{\textbf{Complexity Factor For Anisotropic Source in Non-minimal Coupling Metric $f(R)$ Gravity}}

\author{G. Abbas \thanks{ghulamabbas@iub.edu.pk}
and H. Nazar \thanks{hammadnazar350@gmail.com}
\\Department of Mathematics The Islamia University\\ of Bahawalpur,
Bahawalpur, Pakistan.}
\date{}
\maketitle
\begin{abstract}
In this outline we recognize the idea of complexity factor for static anisotropic self-gravitating source with generalized $f(R)$ metric gravity theory. In present consideration, we express the Einstein field equations, hydrostatic equilibrium equation, the mass function and physical behavior of $f(R)$ model by using some observational data of well known compact stars like $4U~1820-30, SAX~J1808.4-3658$ and $Her~X-1$. We define the scalar functions through the orthogonal splitting of the Reimann-Christofell tensor and then find the vanishing complexity condition for self-gravitating system with the help of these scalars. It has been found that the vanishing condition for the complexity are pressure anisotropy and energy density inhomogeneity must cancel each other. Moreover, we study the momentous results of an astral object for the vanishing of complexity factor. Finally, these solutions reduced to previous investigation about complexity factor in General Relativity by taking $\lambda=0$.
\end{abstract}


\section{Introduction}
In the early 20th century, a remarkable mile stone declared by Albert Einstein on the wonderful
theory of General Relativity (GR) that theory provides the revolutionary break throw in the field of modern science.
Afterwards, many astrophysicists have planed various achievements in the form of modified theories of gravity,
which explain the self-gravitating models, gravitational collapse, gravitating physics, cosmological
physics and relativistic structures. Here now, in this format to need some discussion on highly recent work named as
complexity factor that are strongly debated in self-gravitating relativistic structures with modified gravity theories
and as well as in General Relativity. The word complexity related to this term that have involved ample complications
in a system. Much efforts have been devoted on the well defined description of complexity in the numerous fields of
science \cite{2}-\cite{8}. In this scenario a lot of work have been organized in precise mode that are given in \cite{9}-\cite{15}.

A well defined set note is to be written on the specific definition of complexity that was arranged by Lopez-Ruiz
and co-researcher \cite{8,13,16}. Basically, in diverse situation the definition of the complexity term is directly
associated to the concept of information and entropy that confer the relativistic systems. In the Physics literature
\cite{8} the concluding state of the perfect crystal in which arises the complexity factor and having episodic action
and also has the causal performance of the isolated perfect gas. The perfect crystal is a totaly arranged scheme of atoms
that organized in a systematic style. In this mode to explain the ideal crystal that are described with the slight
necessary part of the information, which gives less complexity in the crystal. On the other hand, the isolated perfect
gas is completely unstable and all the parts have similar contribution to give detail associated to the perfect gas such that
it has a extreme complexity. These structures appear in the basic objects with maximum complexity.

The description of complexity ought to likewise incorporate some different factors past data or direction.
Lopez-Ruiz \cite{8} established the key mechanism of instability which governs distance of the equiprobable
distribution in the structure of available condition. Therefore, in instance of ideal crystal and perfect
gas situation the instability should be extreme and zero respectively. Thus, in originally the complexity
factor obtained from these two concepts such that degree of the disorder of the system (instability) and
information. The picture of instability and information which deals to the probability distribution
is reformulated in \cite{14},\cite{17}-\cite{19} with the term of energy density in the source. Though,
the energy density factor is not sufficient to define the concept of complexity term because the pressure
is not present which discussed in the matter energy-momentum tensor and behavior is significant in the
system formulation of the matter dispersal.

Currently, Herrera \cite{9} examined the self-gravitating source with anisotropic matter distribution,
to find the factor of vanishing complexity by using orthogonal splitting of the curvature tensor
for the structure scalars on the Tolman mass and also studied the general mass function.
Herrera et al. \cite{20} discussed the results of gravitational collapse in context of Israel-Stewart
notion for the viscid dissipative case with heat flux, the bulk and shear viscosity. Herrera et al.
\cite{21} evaluated the solutions for self-relativistic gravitating collapse with spherical symmetric
matter distribution in the dissipative situation undergo heat flow and easy streaming radiation in framework
of Post-quasi static estimation. Herrera and Santos \cite{22} described the study on gravitational collapse
in background of Misner and Sharp approach and extended the work to dissipative case in the form of heat flow
and free radiation streaming, to analyze the dynamical system of equation and then coupled
to heat transportation equations. Herrera et al. \cite{23} concluded the results in self-gravitating collapsing
source with anisotropic configuration for the usual study on such system of equations that bring a general,
essential and appropriate state for the disappearing of the spatial gradients of energy density. Herrera
\cite{24} investigated the effects on spherically symmetric collapsing self-gravitating dense source, to
find the inhomogeneity energy density that comes through the pressure anisotropy and heat dissipation.

Sharif and Zaeem \cite{25} analyzed the consequences in Non-adiabatic case undergoing which possess the heat
flow and free radiation streaming with non-rotating charged plane symmetry and assuming anisotropic source
geometry for the description of structure scalars. Sharif and Yousaf \cite{26} explored the solutions in
cylindrical symmetric model with anisotropic collapsing matter distribution under the expansion-free parameter,
which evaluate the vacuum cavity and also conferred some families solutions and additionally to get each family
have two exact models. Sharif and Zaeem \cite{27} found the impacts on charged cylindrical expansion-free
collapsing source with local anisotropic geometry that established in the significance of disequilibrium condition
under the perturbation approach.

Recently, Abbas and Nazar \cite{10} analyzed the effects on modified $f(R)$ theory of gravity for the
vanishing of complexity of the system by introducing the orthogonal splitting of the Riemann tensor for
the scalar function, in which the source has pressure anisotropy and inhomogeneous energy density that
causes the complexity, when the system have zero complexity then these two factors cancel to each other.
Capozziello \cite{28} considered the Lan$\acute{e}$-Emden equation that comes from $f(R)$ metric theory
and investigated different parameters for the stability of interstellar system. Capozziello \cite{29}
studied the results of modified TOV equation in $f(R)$ metric formalism with the insertion of additional
degree of freedom, describe the effects on mass-radius configuration for static neutron star.
Nojiri and Odintsov \cite{30} examined the impacts on such models that might lead to vital Newton law corrections
at large cosmological scale, that results to describe with cosmological constant for the early and late time accelerating
cosmic evolution in context of $f(R)$ gravity.

Sotiriou and Faraoni \cite{31} studied the significant facts about $f(R)$ gravity theory
with metric-Palatini, metric-Affine approaches and analyze the theoretical features of the gravitating system.
Capozziello \cite{32} discussed the noteworthy issues for cosmic evolution and quintessence
with FRW model in higher order theories of gravity. De Felice and Tsujikawa \cite{33} reviewed
thoroughly several applications of metric theories of gravity for cosmology and gravitation with dark
energy, native gravity limitations, astrophysical perturbations, inflation and spherical
symmetric solutions. Capozziello et al. \cite{34} examined the well known results of matter
and gravitational potential of galaxy cluster projected at X-ray reflection and clarified
without the need of extra dark source in background of modified $f(R)$ theory. Chiba et al. \cite{35}
recognized the rushing cosmic effects with the solar system restrictions in $f(R)$ context.
Nojiri and Odintsov \cite{36} investigated the impacts on such alternative gravity theories
for the unified cosmic picture of early-time inflation with the dark energy epoch.

Bertolami et al. \cite{37} elaborated the results on the system of equation of motion that organized
for the huge particles and also discussed in the presence of extra force that coupled to generic invariant
function $R$ with modified $f(R)$ context. Bertolami et al. \cite{38} studied the key features
of some unlike formations of relativistic isotropic matter Lagrangian densities in non-minimal
coupled matter curvature theory. Garcia and Lobo \cite{39} evaluated the consequences of wormhole structures
for the null energy of normal fluid, which reduces in the framework of non-minimal matter curvature coupling
of $f(R)$ gravity theory. Bertolami and P$\acute{a}$ramos \cite{40} concluded the results of gravitational
non-minimal curvature matter coupling that are associated between metric $f(R)$ formalism and scalar
tensor objects defined with physical metric related to the scalar field and also these solutions confer
on the weak field approximation of $f(R)$ gravity theories. Faraoni \cite{41} extended the work of $f(R)$ gravity
theory with an additional component of Lagrangian matter density that coupled with the usual
Ricci scalar function $R$ which behaves like an extra force that exerted on the system.

Tamanini and Koivisto \cite{42} explored the Lagrangian scalar field to model the
extended non-minimally coupling between matter and curvature theories, which define the supplement force that
comes through non-conservative energy-momentum tensor. Koivisto and Tamanini \cite{43} investigated the noteworthy
issues on non-minimal curvature matter with $f(R)$ gravity theories for the feasible stability states.
Silva and P$\acute{a}$ramos \cite{44} examined the analytical results under the perturbation approach
in self-stable manner for the feasible dark source structure that are expressed in context of modified $f(R)$
gravity theories with non-minimal coupled curvature matter. Branco et al. \cite{45} worked out on
non-minimally coupled matter gravity object in context of perturbation with spherically Minkowskian metric, to
analyze the Starobinsky inflation paradigms and also explain the geodesic effects on the model.
P$\acute{a}$ramos and Ribeiro \cite{46} discussed the feasible dark energy models under the criterion
of dynamical system approach in background of non-minimal curvature coupling theories that define the
useful cosmological consequences.

The order of the profile is as follows: In Sec. \textbf{II} we have designed the spherically symmetric
self-gravitating anisotropic source in non-minimal curvature matter coupling of $f(R)$ gravity theory.
We manipulate their relevant field equations with effective viable results which are used thoroughly in this work and also discuss the physical consequences of $f(R)$ model. The other well specific solutions related to the orthogonal splitting
of the Riemann Christofell tensor and scalar functions have been presented with meaningful sense in Sec. \textbf{III}.
The Sec. \textbf{IV} characterize the exact modified solutions with vanishing of
complexity factor and also discussed the useful relativistic dense object. Finally, in last we present the summary of the paper.

\section{Relativistic System of Equations with Non-Minimal Coupling of $f(R)$ metric Gravity Theory}

Here, we present the interior object formed by spherically symmetric static self-gravitating source that are distinguished
within the framework of non-minimally coupled $f(R)$ gravity theory. The general Einstein-Hilbert action in GR is as follows,
\begin{equation}\label{1}
S_{EH}=\frac{1}{2\kappa}\int d^4x\sqrt{-g}R.
\end{equation}
The generalized action of the matter-curvature with non-minimally coupling in the modified metric $f(R)$ theory
is given by \cite{37,37a}
\begin{equation}\label{2}
S=\int d^4x\sqrt{-g}\left[\frac{1}{2}R+(1+\lambda f(R))L_{m}\right].
\end{equation}
Here $\lambda$ is constant and indicates the coupling strength between matter and geometry, while the
$L_{m}$ is the matter lagrangian density, $f_{i}(R)\quad(i=1, 2)$ are usual functions of curvature scalar $R$ and introduced in above
action with $f_{1}(R)=R$ and $f_{2}(R)=f(R)$, respectively and $8\pi G=c=1$
is used completely in this work. The action (\ref{2}) can be reproduced in context of general relativity by setting $\lambda=0$.

The gravitational field equations for the non-minimal curvature matter coupled of $f(R)$ gravity theory
are obtained by the variation of the action (\ref{2}) with respect to $g_{\mu\nu}$,
\begin{eqnarray}\nonumber
R_{\mu\nu}-\frac{1}{2}R g_{\mu\nu}=&&
-2\lambda f_{R}(R)L_{m}R_{\mu\nu}+2\lambda \left(\nabla_{\mu}\nabla_{\nu}-g_{\mu\nu}\nabla_{\alpha}\nabla^{\alpha}\right)L_{m}f_{R}(R)
\\&&+\left(1+\lambda f(R)\right)T_{\mu\nu}\label{3},
\end{eqnarray}
where $T_{\mu\nu}$, $F_i(R)\equiv f_{iR}(R)$, $\nabla_{\alpha}\nabla^{\alpha}$ and $\nabla_{\mu}$ represent the mass energy-momentum
tensor, the derivative w.r.t. the generic functions of Ricci invariant $R$, the D'Alembertian operator and the general covariant derivative
related with the Levi-Civita connection of the metric, respectively.

Equation (\ref{3}) can be reformulated and taken into the form,
\begin{equation}\label{4}
G_{\mu \nu}=\kappa\Big(T_{\mu \nu}^m+T_{\mu \nu}^{eff}\Big).
\end{equation}
Here $\kappa=\frac{\left(1+\lambda f(R)\right)}{\left(1+2\lambda F(R)L_m\right)}$ is an effective coupling parameter.
\begin{equation}\label{5}
T_{\mu\nu}^{eff}=\frac{1}{(1+\lambda f(R))}\Big[-\lambda R F(R) L_m g_{\mu\nu}
+2\lambda\left(\nabla_{\mu}\nabla_{\nu}-g_{\mu\nu}\nabla^{\alpha}\nabla_{\alpha}\right)F(R)L_m \Big],
\end{equation}
is the effective stress-energy related to the relativistic interior of self-gravitating source in modified $f(R)$ gravity theory with non-minimal coupled curvature and matter.

By considering the given line-element for static spherically symmetric interior geometry is as follows
\begin{equation}\label{6}
ds^{2}=e^{\nu}dt^{2}-e^{\mu}dr^{2}-r^{2}
(d\theta^{2}+\sin^{2}\theta d\phi^{2}),
\end{equation}
where $\nu$ and $\mu$ are such functions that depend on $r$.
The matter energy-momentum tensor for the self-gravitating anisotropic perfect fluid distribution is given below
\begin{equation}\label{20}
T^{\eta}_{\gamma}=\rho v^{\eta}v_{\gamma}-P h^{\eta}_{\gamma}+\Pi^{\eta}_{\gamma}.
\end{equation}
Here $\rho$ is the energy density and the other important conventions are
\begin{eqnarray}\nonumber
&&\Pi^{\eta}_{\gamma}=\Pi(\chi^{\eta}\chi_{\gamma}+\frac{1}{3}h^{\eta}_{\gamma}), \quad P=\frac{1}{3}({P_{r}}+2P_{\bot}),\\&&
\quad \Pi=P_{r}-P_{\bot}, \quad h^{\eta}_{\gamma}=\delta^{\eta}_{\gamma}-v^{\eta}v_{\gamma}.\label{21}
\end{eqnarray}
The non-zero component of four velocity is
\begin{equation}\label{18}
v^{\gamma}=(\frac{1}{e^{\frac{\nu}{2}}},0,0,0),
\end{equation}
where the four vector components are defined under
\begin{equation}\label{22}
\chi^{\gamma}=(0,\frac{1}{e^{\frac{\mu}{2}}},0,0),
\end{equation}
and the following properties are satisfied
\begin{equation}\label{a}
\chi^{\eta}v_{\eta}=0, \quad \chi^{\eta}\chi_{\eta}=-1.
\end{equation}
Now the only non-zero value of four-acceleration from easy to manipulate this term $a_{\beta}=v_{\beta;\gamma} v^{\gamma}$,
\begin{equation}\label{19}
a_{1}=-\frac{1}{2}\nu^{'}.
\end{equation}
The non-vanishing components of the energy-momentum tensor are
\begin{equation}\label{7}
{T^{0}_{0}}=\rho,
\end{equation}
\begin{equation}\label{8}
{T^{1}_{1}}=-P_{r},
\end{equation}
\begin{equation}\label{9}
{T^{2}_{2}}={T^{3}_{3}}=-P_{\bot}.
\end{equation}
Now Eq.(\ref{3}) gives the system of equations in modified $f(R)$ gravity theory with non-minimally coupled to matter curvature is,
\begin{eqnarray}\nonumber
-\left(e^{-\mu}\left(\frac{1}{r^2}-\frac{\mu'}{r}\right)-\frac{1}{r^2}\right)&&=\kappa\Big[\rho+\frac{1}{(1+\lambda f(R))}\Big\{-\lambda R F L_m
+2\lambda e^{-\mu}\Big(F L_m \Big)''\\&&+2\lambda e^{-\mu}\left(\frac{2}{r}-\frac{\mu'}{r}\right)\Big(F L_m\Big)'\Big\}\Big]\label{10},\\
\left(e^{-\mu}\left(\frac{1}{r^2}+\frac{\nu^{'}}{r}\right)-\frac{1}{r^2}\right)&&=\kappa\Big[P_r-\frac{1}{(1+\lambda f(R))}\Big\{-\lambda R F L_m
+2\lambda e^{-\mu}\Big(\frac{\nu'}{2}+\frac{2}{r}\Big)\Big(F L_m\Big)'\Big\}\Big],\nonumber\\\label{11}\\
\frac{e^{-\mu}}{4}\Big(2\nu''+\nu'^2-\mu'\nu'+2\frac{\nu'-\mu'}{r}\Big)&&=\kappa\Big[P_{\bot}-\frac{1}{(1+\lambda f(R))}\Big\{-\lambda R F L_m
+2\lambda e^{-\mu}\Big(F L_m\Big)''\nonumber\\&&+2\lambda e^{-\mu}\Big(\frac{\nu'}{2}-\frac{\mu'}{2}+\frac{1}{r}\Big)\Big(F L_m\Big)'\Big\}\Big]\label{12}.
\end{eqnarray}
Here prime sign denotes the partial derivative w.r.t. $r$. The hydrostatic equilibrium equation can be
obtained from generalized conservation equation
\begin{equation}\label{13}
P'_{r}=-\frac{(\rho+P_{r})}{2}\nu'+2\frac{(P_{\bot}-P_{r})}{r}-\frac{\lambda F}{(1+\lambda f(R))}[L_m+P_r]R'.
\end{equation}
It can also be recognized as the standardized Tolman-Oppenheimer-Volkoff equation for anisotropic self-gravitating source
in non-minimal modified $f(R)$ gravity theory.

We use Eq.(\ref{11}), and obtained that
\begin{equation}\label{14}
\nu^{'}=2\frac{\Big(m+\frac{\kappa P_{r}r^3}{2}\Big)}{r(r-2m)}-\frac{\kappa r^3}{r(r-2m)(1+\lambda f(R))}\Big\{-\lambda R F L_m+2\lambda e^{-\mu}\Big(\frac{\nu'}{2}+\frac{2}{r}\Big)\Big(F L_m\Big)'\Big\}.
\end{equation}
It can be written as of Eq.(\ref{13})
\begin{eqnarray}\nonumber
&&P'_{r}=-\frac{\Big(m+\frac{\kappa P_{r}r^3}{2}\Big)(\rho+P_{r})}{r(r-2m)}+\frac{\kappa(\rho+P_{r})r^2}{2(r-2m)(1+\lambda f(R))}\Big\{-\lambda R F L_m\\&&+2\lambda e^{-\mu}\Big(\frac{\nu'}{2}+\frac{2}{r}\Big)\Big(F L_m\Big)'\Big\}+2\frac{(P_{\bot}-P_{r})}{r}-\frac{\lambda F}{(1+\lambda f(R))}[L_m+P_r]R'\label{15}.
\end{eqnarray}
The mass function $m(r)$ is defined by the following relation \footnote{ which is same as defined by Eq.(9) in Astashenok et al. \cite{47}}
\begin{equation}\label{16}
m=\frac{r(1-e^{-\mu})}{2},
\end{equation}
we use Eq.(\ref{10}) with the above information and obtained,
\begin{eqnarray}\nonumber
m=&&\frac{1}{2}\int^{r}_{0}\kappa \rho \tilde{r}^2 d\tilde{r}+\frac{1}{2}\int^{r}_{0}\frac{\kappa\tilde{r}^2}{(1+\lambda f(R))}\Big[-\lambda R F L_m+e^{-\mu}
\Big\{2\lambda (F L_m)''\\&&+2\lambda \Big(\frac{2}{r}-\frac{\mu^{'}}{2}\Big)(F L_m)'\Big\}\Big]d\tilde{r}.\label{17}
\end{eqnarray}
\subsection{The Conformal curvature tensor}
Now, we define the Weyl tensor, which comprises into two parts one is the magnetic
part and the other is electric part. The magnetic part vanishes due to spherical symmetric system
while the other electric part has taken into the form
\begin{equation}\label{23}
E_{\phi\sigma}=C_{\phi\psi\sigma\xi}v^{\psi}v^{\xi},
\end{equation}
where
\begin{eqnarray}
C_{\mu\nu\alpha\beta}=(g_{\mu\nu\kappa\lambda}g_{\alpha\beta\psi\xi}-\eta_{\mu\nu\kappa\lambda}\eta_{\alpha\beta\psi\xi})v^{\kappa}v^{\psi}E^{\lambda\xi}.\label{24}
\end{eqnarray}
Here $ g_{\mu\nu\kappa\lambda}= g_{\mu\kappa}g_{\nu\lambda}-g_{\mu\lambda}g_{\nu\kappa}$, and $\eta_{\mu\nu\kappa\lambda}$ is the Levi-Civita tensor.
It is the rewrite form of the $E_{\kappa\lambda}$, as given below
\begin{equation}\label{25}
E_{\kappa\lambda}=E(\chi_{\kappa}\chi_{\lambda}+\frac{1}{3}h_{\kappa\lambda}),
\end{equation}
with
\begin{equation}\label{26}
E=-\frac{1}{4e^{\mu}}\left[\nu^{''}+\frac{\nu^{'2}}{2}-\frac{\mu'\nu^{'}}{2}-\frac{\nu^{'}}{r}+\frac{\mu'}{r}+\frac{2}{r^2}-\frac{2e^{\mu}}{r^2}\right],
\end{equation}
and the following arrangements to complete the conditions
\begin{equation}\label{27}
E^{\kappa}_{\kappa}=0, \quad E_{\kappa\psi}=E_{(\kappa\psi)}, \quad E_{\kappa\psi}v^{\psi}=0.
\end{equation}
\subsection{Profile of general mass function and Tolman mass }
Now in this format, we define two basic functions of mass usually known as General mass and the Tolman mass
that described for the interior boundary of the relativistic source and its physical significance connected with Weyl tensor.
Furthermore, next we will use for the validation of the complexity factor.

In the association of Eqs.(\ref{10})-(\ref{12}) and (\ref{16}), we get
\begin{eqnarray}
m=\frac{\kappa r^3(\rho+P_\bot-P_r)}{6}+\frac{1}{3}Er^3+\frac{\kappa r^3}{6(1+\lambda f(R))}\Big\{-\lambda R F L_m+\frac{6\lambda (FL_m)'}{re^{\mu}}\Big\}\label{28}.
\end{eqnarray}
We can rewrite as under
\begin{eqnarray}\nonumber
&&E=-\frac{1}{2r^3}\int^{r}_{0}\tilde{r}^3\Big(\kappa{'}\rho+\kappa \rho{'}\Big)d\tilde{r}+\frac{1}{2}\int^{r}_{0}\frac{\kappa\tilde{r}^2}{(1+\lambda f(R))}\Big[-\lambda R F L_m+2\lambda e^{-\mu}\Big\{(F L_m)''\\&&+ \Big(\frac{2}{\tilde{r}}-\frac{\mu^{'}}{2}\Big)(F L_m)'\Big\}\Big]d\tilde{r}+\frac{\kappa(P_{r}-P_{\bot})}{2}-\frac{\kappa}{2(1+\lambda f(R))}\Big\{-\lambda R F L_m+\frac{6\lambda (F L_m)'}{re^{\mu}}\Big\}\label{29},
\end{eqnarray}
consequently, we use Eq.(\ref{29}) in Eq.(\ref{28}), and can get
\begin{eqnarray}\nonumber
&&m(r)=\frac{\kappa r^3  \rho }{6}-\frac{1}{6}\int^{r}_{0}\tilde{r}^3\Big(\kappa{'}\rho+\kappa \rho{'}\Big)d\tilde{r}+\frac{r^3}{6}\int^{r}_{0}\frac{\kappa\tilde{r}^2}{(1+\lambda f(R))}\Big[-\lambda R F L_m+2\lambda e^{-\mu}\Big\{(F L_m)''\\&&+ \Big(\frac{2}{\tilde{r}}-\frac{\mu^{'}}{2}\Big)(F L_m)'\Big\}\Big]d\tilde{r}\label{30}.
\end{eqnarray}
In the significance of Eq.(\ref{29}), we found that $E$ remarks two physical quantities called inhomogeneous energy density and pressure
anisotropy of the self-gravitating fluid distribution and Eq.(\ref{30}) is related to the mass function that express in case of
energy density homogeneity distribution and the variation introduced by inhomogeneous density in context of
non-minimal coupling to curvature-matter with $f(R)$ theory of gravity.

Alternatively, a momentous definition of energy content proposed by Tolman \cite{48} for static spherical
structure is given by
\begin{equation}\label{31}
m_{T}=\frac{1}{2}\int^{r\Sigma}_{0}\kappa r^2e^\frac{(\nu+\mu)}{2}(\rho+P_r+2P_{\bot})dr.
\end{equation}
The whole energy of the self-gravitating fluid inside the relativistic spherical object of radius $r$ is
\begin{equation}\label{32}
m_{T}=\frac{1}{2}\int^{r}_{0}\kappa\tilde{r}^2 e^\frac{(\nu+\mu)}{2}(\rho+P_r+2P_{\bot})d\tilde{r}.
\end{equation}
The familiar role of $m_{T}$ known as active gravitational mass that played a vital insertion for the expandable global
concept of energy to a local level. The sufficient backup is formulated in \cite{12, 49, 50}
\begin{eqnarray}\nonumber
&&m_{T}=e^\frac{(\nu+\mu)}{2}\Big[m(r)+\frac{\kappa P_{r}r^3}{2}-\frac{\kappa r^3}{2(1+\lambda f(R))}\Big\{-\lambda R F L_m+ \frac{2\lambda}{e^{\mu}}\Big(\frac{\nu^{'}}{2}+
\frac{2}{r}\Big)\Big(FL_m\Big)'\Big\}\Big]\\&&+\int^{r}_{0}\frac{\kappa\tilde{r}^2e^{(\frac{\nu+\mu}{2})}}{(1+\lambda f(R))}\Big[-\lambda R F L_m+\frac{\lambda}{e^{\mu}}\Big\{(FL_m)''-
\Big(\mu'-3\nu^{'}\Big)\frac{(FL_m)'}{2}+\frac{2}{\tilde{r}}\Big(FL_m\Big)'\Big\}\Big]d\tilde{r}.\label{33}
\end{eqnarray}
With the help of Eq.(\ref{14}), it becomes
\begin{eqnarray}\nonumber
&&m_{T}=\frac{e^\frac{(\nu-\mu)}{2}\nu^{'}r^2}{2}+\int^{r}_{0}\frac{\kappa\tilde{r}^2e^{(\frac{\nu+\mu}{2})}}{(1+\lambda f(R))}\Big[-\lambda R F L_m+\frac{\lambda}{e^{\mu}}\Big\{(FL_m)''-
\Big(\mu'-3\nu^{'}\Big)\frac{(FL_m)'}{2}\\&&+\frac{2}{\tilde{r}}\Big(FL_m\Big)'\Big\}\Big]d\tilde{r}.\label{34}
\end{eqnarray}
The worth of Eq.(\ref{34}) can also be recognized as active gravitational mass of the structure.
By using Eq.(\ref{19}) and replacing into $(a=-\psi^{\alpha}a_{\alpha})$, for the gravitational acceleration of test particles and can get
\begin{eqnarray}\nonumber
&&a=\frac{1}{e^\frac{\nu}{2}r^2}\Big[m_{T}-\int^{r}_{0}\frac{\kappa\tilde{r}^2e^{(\frac{\nu+\mu}{2})}}{(1+\lambda f(R))}\Big[-\lambda R F L_m+\frac{\lambda}{e^{\mu}}\Big\{(FL_m)''-
\Big(\mu'-3\nu^{'}\Big)\frac{(FL_m)'}{2}\\&&+\frac{2}{\tilde{r}}\Big(FL_m\Big)'\Big\}\Big]d\tilde{r}.\label{35}
\end{eqnarray}
We differentiate Eq.(\ref{34}) w.r.t $r$, then use system of equations and associated with (\ref{33}), it turned out that
\begin{eqnarray}\nonumber
&&r m'_{T}-3m_{T}=e^\frac{(\nu+\mu)}{2}r^3\Big[\frac{\kappa(P_{\bot}-P_r)}{2}-E-\frac{\kappa\lambda}{(1+\lambda f(R))e^{\mu}}\Big\{(FL_m)''-\Big(\frac{\mu'}{2}+\frac{1}{r}\Big)\Big(FL_m\Big)'\Big\}\\\nonumber&&+\frac{\kappa}{(1+\lambda f(R))}\Big\{-\lambda R F L_m+\frac{\lambda}{e^{\mu}}\Big((FL_m)''-
\Big(\mu'-3\nu^{'}\Big)\frac{(FL_m)'}{2}+\frac{2}{r}\Big(FL_m\Big)'\Big)\Big\}\Big]\\&&-3\int^{r}_{0}\frac{\kappa\tilde{r}^2e^{(\frac{\nu+\mu}{2})}}
{(1+\lambda f(R))}\Big[-\lambda R F L_m+\frac{\lambda}{e^{\mu}}\Big\{(FL_m)''-
\Big(\mu'-3\nu^{'}\Big)\frac{(FL_m)'}{2}+\frac{2}{\tilde{r}}\Big(FL_m\Big)'\Big\}\Big]d\tilde{r}.\label{36}
\end{eqnarray}
In the integral form, it is easy to manipulate that
\begin{eqnarray}\nonumber
&&m_{T}=(m_{T})_{\Sigma}\left(\frac{r}{r_{\Sigma}}\right)^3-r^3\int^{r_{\Sigma}}_{r}\frac{e^\frac{(\nu+\mu)}{2}}{\tilde{r}}
\Big[\frac{\kappa(P_{\bot}-P_r)}{2}-E-\frac{\kappa\lambda}{(1+\lambda f(R))e^{\mu}}\Big\{(FL_m)''\\\nonumber&&-\Big(\frac{\mu'}{2}+\frac{1}{\tilde{r}}\Big)\Big(FL_m\Big)'\Big\}+\frac{\kappa}{(1+\lambda f(R))}\Big\{-\lambda R F L_m+\frac{\lambda}{e^{\mu}}\Big((FL_m)''-
\Big(\mu'-3\nu^{'}\Big)\frac{(FL_m)'}{2}+\frac{2}{\tilde{r}}\Big(FL_m\Big)'\Big)\Big\}\Big]d\tilde{r}\\&&
+3r^3\int^{r_{\Sigma}}_{r}\Big(\frac{1}{\tilde{r}^4}\int^{r}_{0}\frac{\kappa\tilde{r}^2e^{(\frac{\nu+\mu}{2})}}
{(1+\lambda f(R))}\Big[-\lambda R F L_m+\frac{\lambda}{e^{\mu}}\Big\{(FL_m)''-
\Big(\mu'-3\nu^{'}\Big)\frac{(FL_m)'}{2}+\frac{2}{\tilde{r}}\Big(FL_m\Big)'\Big\}\Big]d\tilde{r}\Big)d\tilde{r}.\label{37}
\end{eqnarray}
Here for the assistance of Eq.(\ref{29}), one can get
\begin{eqnarray}\nonumber
&&m_{T}=(m_{T})_{\Sigma}\left(\frac{r}{r_{\Sigma}}\right)^3-r^3\int^{r_{\Sigma}}_{r}e^\frac{(\nu+\mu)}{2}
\Big[\frac{\kappa(P_{\bot}-P_r)}{\tilde{r}}+\frac{1}{2\tilde{r}^4}\int^{r}_{0}\tilde{r}^3\Big(\kappa{'}\rho+\kappa\rho{'}\Big)d\tilde{r}\\\nonumber&&
-\frac{\kappa\lambda}{\tilde{r}(1+\lambda f(R))e^{\mu}}\Big\{(FL_m)''-\Big(\frac{\mu'}{2}+\frac{1}{\tilde{r}}\Big)\Big(FL_m\Big)'\Big\}+\frac{\kappa}{\tilde{r}(1+\lambda f(R))}\Big\{-\frac{3}{2}\lambda R F L_m+\frac{\lambda}{e^{\mu}}\Big((FL_m)''\\\nonumber&&-
\Big(\mu'-3\nu^{'}\Big)\frac{(FL_m)'}{2}+\frac{5}{\tilde{r}}\Big(FL_m\Big)'\Big)\Big\}-\frac{1}{2\tilde{r}}\int^{r}_{0}\frac{\kappa\tilde{r}^2}{(1+\lambda f(R))}\Big[-\lambda R F L_m+2\lambda e^{-\mu}\Big\{(F L_m)''\\\nonumber&&+ \Big(\frac{2}{\tilde{r}}-\frac{\mu^{'}}{2}\Big)(F L_m)'\Big\}\Big]d\tilde{r}\Big]d\tilde{r}
+3r^3\int^{r_{\Sigma}}_{r}\Big(\frac{1}{\tilde{r^4}}\int^{r}_{0}\frac{\kappa\tilde{r}^2e^{(\frac{\nu+\mu}{2})}}{(1+\lambda f(R))}\Big[-\lambda R F L_m+\frac{\lambda}{e^{\mu}}\Big\{(FL_m)''-\Big(\mu'-3\nu^{'}\Big)\\&&\frac{(FL_m)'}{2}+\frac{2}{\tilde{r}}
\Big(FL_m\Big)'\Big\}\Big]d\tilde{r}\Big)d\tilde{r}.\label{38}
\end{eqnarray}
The main objective of the above expression is to notify the ample effects of inhomogeneous energy density and pressure anisotropy arrived in the second
integral onto the Tolman's mass in the framework of non-minimal coupled to matter curvature with modified $f(R)$ metric gravity theory. Nowhere, in this interpretation we would like to indicate that when we take $\lambda=0$, all the settings reproduce into Herrera \cite{9}.
\subsection{Analysis of $f(R)$ Model and Observational Data}

In order to discuss the physical nature of formulated results, we have to chose the particular form of $f(R)$ non-minimal coupled model, for the sake of simplicity we take $f(R)=e^{\alpha R}$, in which the exponential $f(R)$ non-minimal coupled model, where $\alpha$ is free parameter.\textbf{ This is an exact feasible $f(R)$ model for the sake of convenient results in the dark source case. It has been established by Cognola et al. \cite{50aa} that a class of exponential theories of $f(R)$ gravity can be introduced which are capable of unifying the dark energy with inflation. The exponential models proposed by the authors are important for phenomenological aspects and provides natural way to check viability of modified theories of gravity. Further, all these models passes local test, nonviolation of Newton's law and production of heavy mass for the additional scalar degree of freedom.}

\begin{figure}
\begin{center}
\includegraphics[width=.44\linewidth, height=2in]{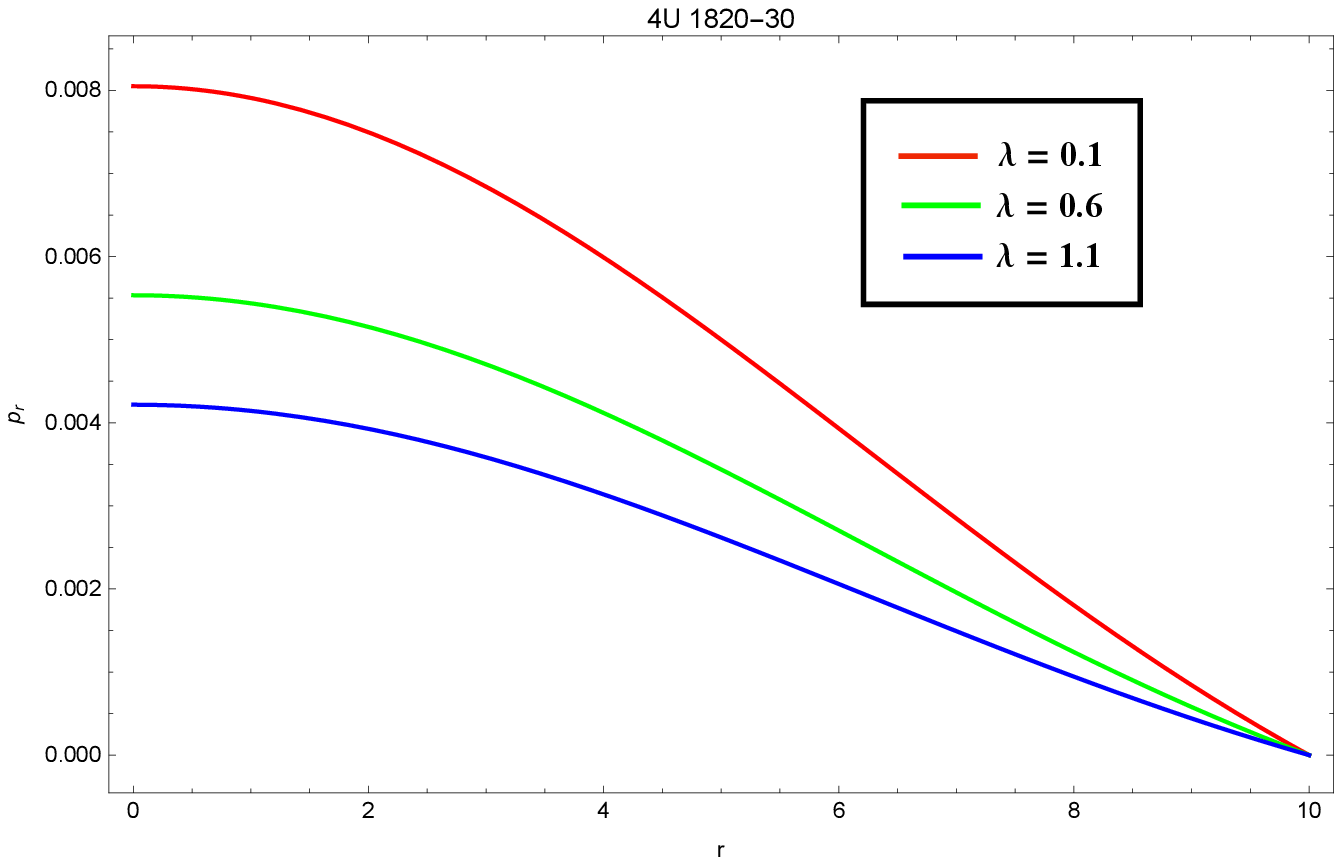}
\includegraphics[width=.44\linewidth, height=2in]{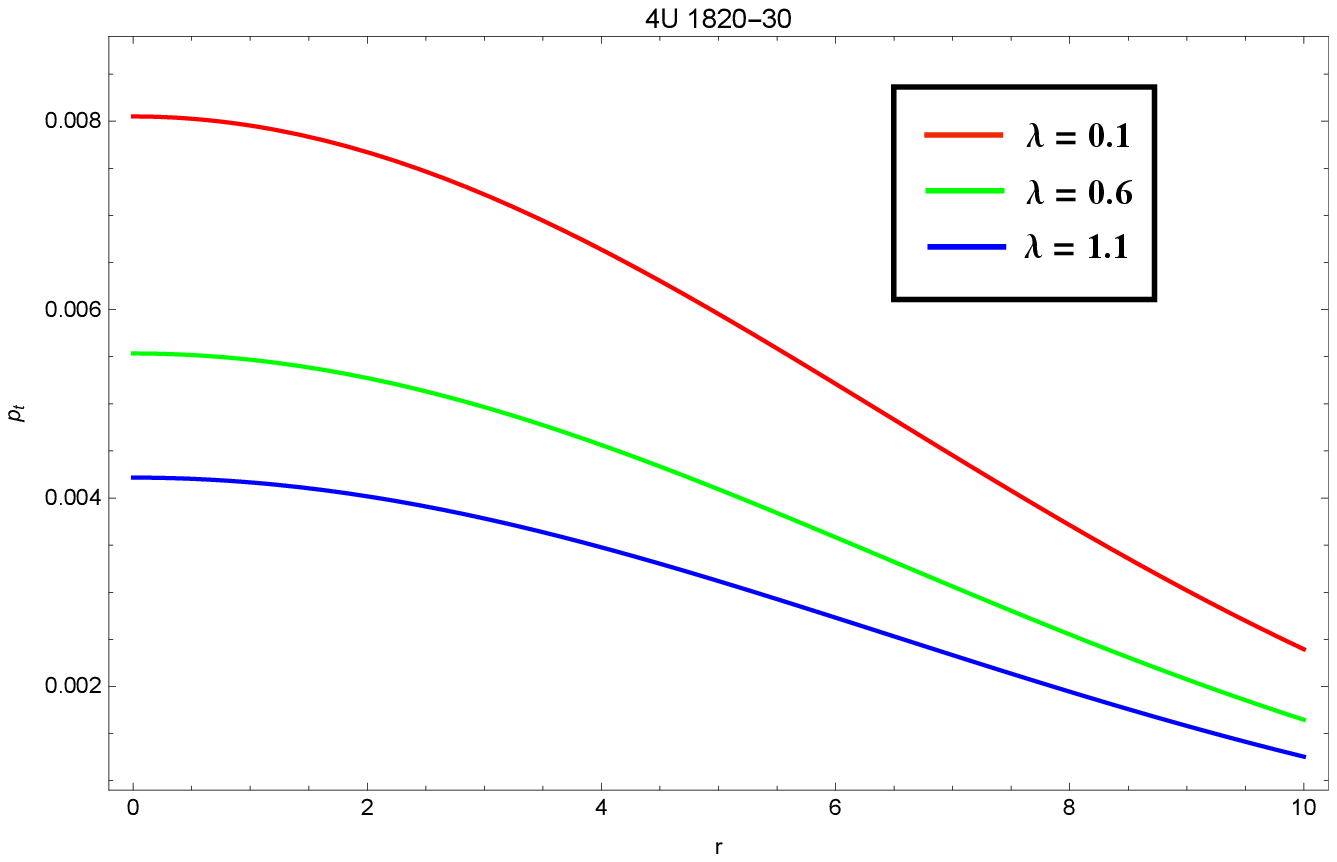}
\caption{Variation of $P_{r}$ (left panel) and $P_{t}$ (right panel). For graphical representation, we take $\alpha=1.1$ and $a=1.02$. }
\end{center}
\end{figure}

\begin{figure}
\begin{center}
\includegraphics[width=.44\linewidth, height=2in]{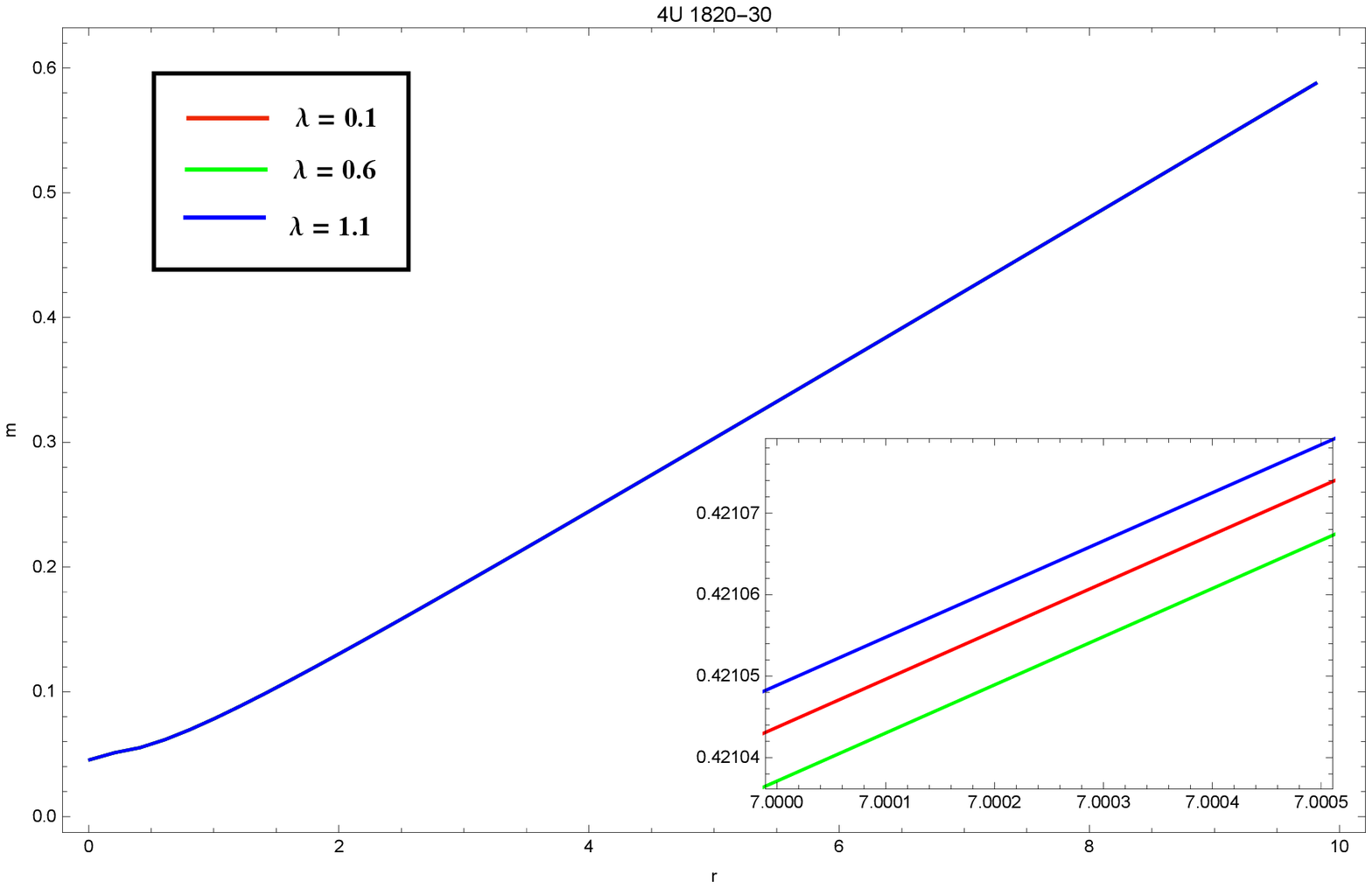}
\includegraphics[width=.44\linewidth, height=2in]{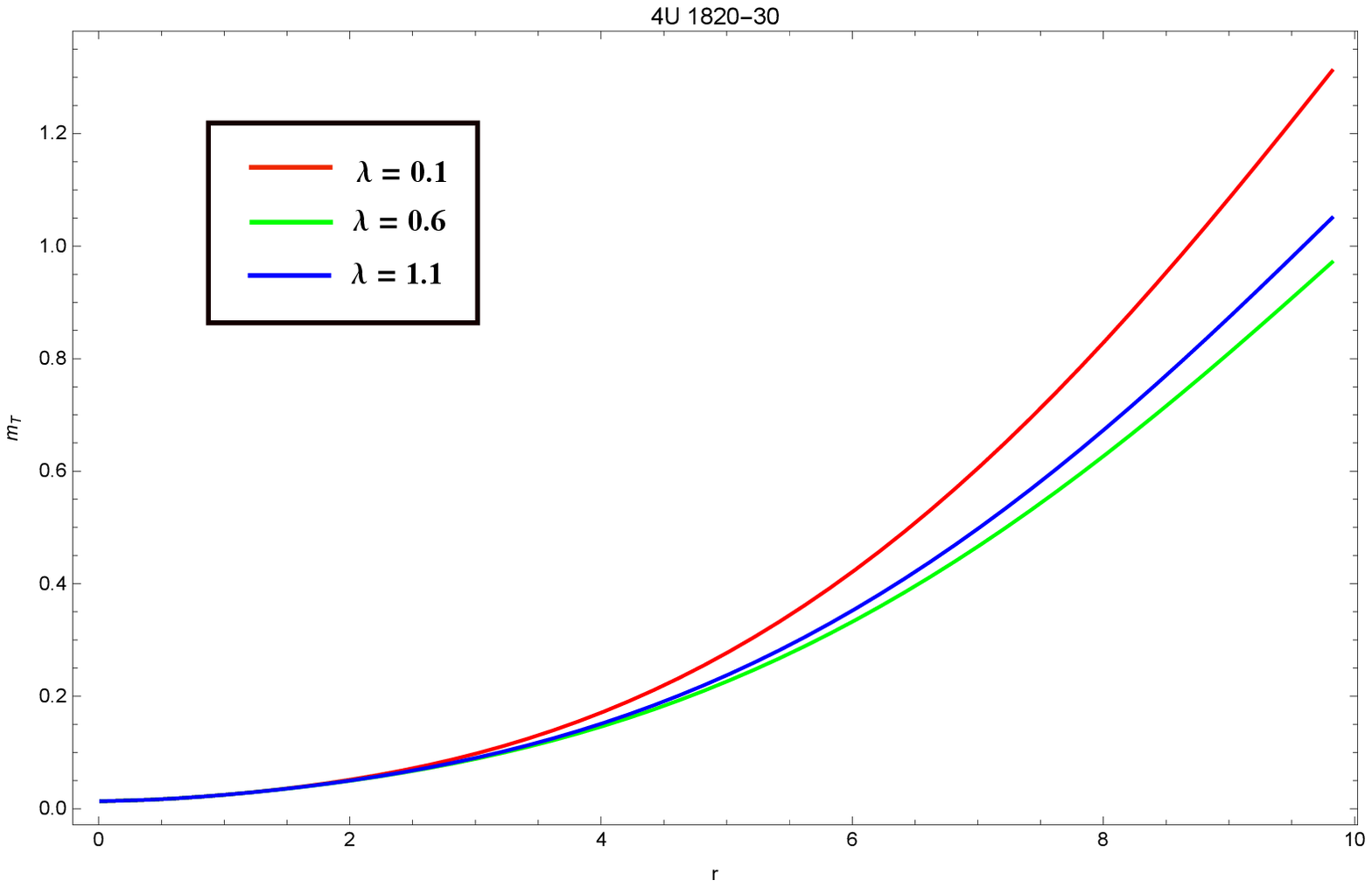}
\caption{Behavior of mass function (left panel) and Tolman mass function (right panel). }
\end{center}
\end{figure}

\begin{figure}
\begin{center}
\includegraphics[width=.44\linewidth, height=2in]{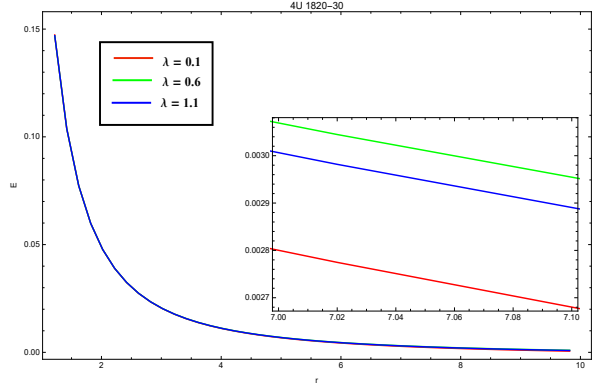}
\caption{Energy profile }
\end{center}
\end{figure}

Moreover, in our framework we choose a simple class models for the spatial astral density, which in idea resemble de Vaucoulour's account in the exterior regions, but not definitely in the center. These two class of models were firstly introduced by Jaffe \cite{50a} and Hernquist \cite{50b} which have central astral densities proportional to $r^{-2}$ and $r^{-1}$, such two type of models can be induced to a family of density profiles with diverse central slopes given by
\begin{equation}\label{38a}
\rho(r)=\frac{(3-\gamma)M a}{4\pi r^{\gamma}(a+r)^{4-\gamma}}.
\end{equation}
Here $M$ is a whole mass and $a$ is a scaling radius. The mass distribution in the center are proportional to $r^{-\gamma}$, hence the models
by Jaffe and Hernquist are compatible with $\gamma=2$ and $1$, respectively. The $\gamma$ value is confined to the interval $[0,3)$.  In our case we choose the Hernquist model for $\gamma=1$. The metric functions have been chosen in the form of Krori-Barua ansatz \cite{50c} such as $\mu=Ar^2$ and $\nu=Br^2+C$, now by applying matching conditions between interior metric (\ref{6}) and exterior metric which is taken as Schwarzschild metric is given by
\begin{equation}\label{38b}
 ds^2=\left(1-\frac{2M}{r}\right)dt^2-\left(1-
 \frac{2M}{r}\right)^{-1}dr^2-r^2(d\theta^2+sin^2{\theta}d\phi^2).
\end{equation}
The continuity of line elements at $r=R$ gives
\begin{eqnarray}\label{38c}
  g_{tt}^-=g_{tt}^+,~~~~~
   g_{rr}^-=g_{rr}^+,~~~~~
   \frac{\partial g_{tt}^-}{\partial r}=\frac{\partial g_{tt}^+}{\partial r},
  \end{eqnarray}
where $-$ and $+$ signs represent the interior and exterior surface of the star, thus we obatain
 \begin{eqnarray}\label{38d}
  A&=&-\frac{1}{R^2}ln\left(1-\frac{2M}{R}\right),\\\label{24}
 B&=&\frac{M}{R^3}{{\left(1-\frac{2M}{R}\right)}^{-1}},\\\label{24a}
 C&=&ln\left(1-\frac{2M}{R}\right)-\frac{M}{R}{{\left(1-\frac{2M}{R}\right)}^{-1}}.
\end{eqnarray}
Lie et al. \cite{50d} have analyzed the mass-radius connection of X-ray pulsar SAX J1808.4-3658 with two
candidates such as compact star and neutron star mass-radius connection and also described the
constancy of strange star model with SAX J1808.4-3658. They proposed that SAX J1808.4-3658
is a possible strange star model and calculated radii and masses of strange star as $7.07km$, $6.53km$
and  $1.44M_\odot$,$1.32M_\odot$, respectively. Zhang et al. \cite{50e} reviewed 4U 1820-30 for the mass size of
neutron star and testified mass of the direction  $\simeq2.2M\odot$. Guver et al. \cite{50f} investigated 4U 1820-30
for measuring the radius and mass of neutron star with  $1\sigma$ error as $R=9.11\pm0.4km$ and a mass of
$M =1.58\pm0.06M_\odot$. Though, upper bound limit in this calculation is invariant with that in the  Zhang et al
paper. Actually, there is a firm doubt in size of radius and mass of a dense stars. Abubekerov et al. \cite{50g} determined
the mass of Her X-1 using more current and physically acceptable methods and create two unlike estimations of masses
$m_x=0.85\pm0.15M_\odot$ and $m_x=1.8M_\odot$ through the radial velocity curves. This doubt may be due to the rigid
X-ray heating in Her X-1. Few of the researchers \cite{50d,50h} adopted those observed values of $R$ and $M$ for numerous dense stars, now the
table \textbf{1} contained values of $A$, $B$ and $C$ that obtained from Krori-Barua ansatz.

In Fig. \textbf{1}, we explain the graphical representation of viable $f(R)=e^{\alpha R}$ model to the numerical values of $P_{r}$ (Left) and $P_{t}$ (Right) functions for different values of $\lambda$. It can be seen from Fig. \textbf{1} that $P_{r}$ graph showing the gradual decreasing nature from center to the surface of the compact object and vanishes at the surface. Similarly, transverse pressure is maximum at the center and declining towards the surface. Fig. \textbf{2} illustrates the graphical significance on the general mass function (Left) and Tolman mass function (Right) that represents the effective behaviors for the $f(R)$ dark source model. One can see from Fig.  \textbf{2} (Left), the clear image of the combined behavior of mass function reflects through magnification of the small patch of the mass plot. The behavior of the mass function steadily increasing along the radius of the compact star. In case of Tolman mass, the graphical behavior shows the increasing nature from center to surface of the compact object. At the end, both graphs of the mass functions showing very useful results for the significance of viable $f(R)$ model.

In Fig. \textbf{3}, we see the graphical behavior of energy function for the feasible $f(R)$ non-minimal coupled model, in which it is explained that
either the value of energy function increased or decreased at that point. We can see from the plot of energy function that energy is extremely increasing
along with center and decreasing towards the surface. In this stance, the magnified image of the energy plot has presented the better graphical behavior to the feasible $f(R)$ model.
\begin{table}[ht]
\caption{Values of $A$, $B$ and $C$}
\begin{center}
\begin{tabular}{|c|c|c|c|c|c|}
\hline {Compact Star}&  \textbf{ $M$} & \textbf{$R(km)$} &\textbf{ $A(km ^{-2})$}& \textbf{$B(km ^{-2})$} & \textbf{$C$}
\\\hline $4U 1820-30$&2.25$M_\odot$& 10.0 &0.010906441192 &
$0.0098809523811$ & $-2.0787393571141$
\\\hline $SAX J 1808.4-3658$& 1.435$M_\odot$& 7.07& 0.018231569740 &
$0.014880115692$ & $-1.6550842848905$
\\\hline  $Her X-1$& 0.88$M_\odot$& 7.7& 0.0069062764281 &
$0.0042673646183$ & $-0.6624851776985$
\\\hline
\end{tabular}
\end{center}
\end{table}

\section{Scalar Structures }
In this conjecture, we use orthogonal splitting of the curvature tensor familiarized by Bel \cite{51}
and get scalar functions which might be helpful for the vanishing of complexity factor.
The following tensors are introduced by \cite{12,51} for the orthogonal splitting of the curvature tensor,
\begin{equation}\label{39}
Y_{\phi\sigma}=R_{\phi\psi\sigma\xi}v^{\psi}v^{\xi},
\end{equation}
\begin{equation}\label{40}
X_{\phi\sigma}=^*R^{*}_{\phi\psi\sigma\xi}v^{\psi}v^{\xi}=\frac{1}{2}\eta^{\tau\nu}_{\phi\psi}R^{*}_{\tau\nu\sigma\xi}v^{\psi}v^{\xi}.
\end{equation}

Here $R^{*}_{\phi\sigma\psi\xi}=\frac{1}{2}\eta_{\tau\nu\psi\xi}R^{\tau\nu}_{\phi\sigma}$.
Through above expressions, we can define two explicit tensors in the form of
$Y_{\alpha\beta}$ and $X_{\alpha\beta}$ mentioned under:
\begin{equation}\label{41}
Y_{\phi\sigma}=\frac{(\rho+3P)h_{\phi\sigma}}{6}+\frac{\Pi_{\phi\sigma}}{2}+E_{\phi\sigma},
\end{equation}
\begin{equation}\label{42}
X_{\phi\sigma}=\frac{\rho h_{\phi\sigma}}{3}+\frac{\Pi_{\phi\sigma}}{2}-E_{\phi\sigma}.
\end{equation}
We can obtain these four structure scalars $X_{T},X_{TF},Y_{T},Y_{TF}$ in the form of trace-free and trace parts as follows \cite{12}
\begin{equation}\label{43}
X_{T}=\rho,
\end{equation}
\begin{equation}\label{44}
X_{TF}=\frac{\Pi}{2}-E,
\end{equation}
applying Eq.(\ref{29}) and becomes,
\begin{eqnarray}\nonumber
&&X_{TF}=\frac{(P_{r}-P_{\bot})}{2}+\frac{1}{2r^3}\int^{r}_{0}\tilde{r}^3\Big(\kappa{'}\rho+\kappa \rho{'}\Big)d\tilde{r}-\frac{1}{2}\int^{r}_{0}\frac{\kappa\tilde{r}^2}{(1+\lambda f(R))}\Big[-\lambda R F L_m\\\nonumber&&+2\lambda e^{-\mu}\Big\{(F L_m)''+ \Big(\frac{2}{\tilde{r}}-\frac{\mu^{'}}{2}\Big)(F L_m)'\Big\}\Big]d\tilde{r}+\frac{\kappa(P_{\bot}-P_{r})}{2}+\frac{\kappa}{2(1+\lambda f(R))}\Big\{-\lambda R F L_m\\&&+\frac{6\lambda (F L_m)'}{re^{\mu}}\Big\}\label{45},
\end{eqnarray}
\begin{equation}\label{46}
Y_{T}=\frac{1}{2}(\rho+3P_{r}-2\Pi),
\end{equation}
\begin{equation}\label{47}
Y_{TF}=\frac{\Pi}{2}+E,
\end{equation}
by the information of Eq.(\ref{29})
\begin{eqnarray}\nonumber
&&Y_{TF}=\frac{(P_{r}-P_{\bot})}{2}-\frac{1}{2r^3}\int^{r}_{0}\tilde{r}^3\Big(\kappa{'}\rho+\kappa \rho{'}\Big)d\tilde{r}+\frac{1}{2}\int^{r}_{0}\frac{\kappa\tilde{r}^2}{(1+\lambda f(R))}\Big[-\lambda R F L_m\\\nonumber&&+2\lambda e^{-\mu}\Big\{(F L_m)''+ \Big(\frac{2}{\tilde{r}}-\frac{\mu^{'}}{2}\Big)(F L_m)'\Big\}\Big]d\tilde{r}-\frac{\kappa(P_{\bot}-P_{r})}{2}-\frac{\kappa}{2(1+\lambda f(R))}\Big\{-\lambda R F L_m\\&&+\frac{6\lambda (F L_m)'}{re^{\mu}}\Big\}.\label{48}
\end{eqnarray}
From the overhead results of $X_{TF}$ and $Y_{TF}$, we govern the pressure anisotropy
\begin{equation}\label{49}
X_{TF}+Y_{TF}=\Pi,
\end{equation}
The physical significance of $Y_{T}$ and $Y_{TF}$ are getting through the concluding results of Eq.(\ref{47}) used in Eq.(\ref{37}) and
taken into the form
\begin{eqnarray}\nonumber
&&m_{T}=(m_{T})_{\Sigma}\left(\frac{r}{r_{\Sigma}}\right)^3+r^3\int^{r_{\Sigma}}_{r}\frac{e^\frac{(\nu+\mu)}{2}}{\tilde{r}}
\Big[\kappa Y_{TF}+E(1-\kappa)+\frac{\kappa\lambda}{(1+\lambda f(R))e^{\mu}}\Big\{(FL_m)''\\\nonumber&&-\Big(\frac{\mu'}{2}+\frac{1}{\tilde{r}}\Big)\Big(FL_m\Big)'\Big\}-\frac{\kappa}{(1+\lambda f(R))}\Big\{-\lambda R F L_m+\frac{\lambda}{e^{\mu}}\Big((FL_m)''-
\Big(\mu'-3\nu^{'}\Big)\frac{(FL_m)'}{2}+\frac{2}{\tilde{r}}\Big(FL_m\Big)'\Big)\Big\}\Big]d\tilde{r}\\&&
+3r^3\int^{r_{\Sigma}}_{r}\Big(\frac{1}{\tilde{r}^4}\int^{r}_{0}\frac{\kappa\tilde{r}^2e^{(\frac{\nu+\mu}{2})}}
{(1+\lambda f(R))}\Big[-\lambda R F L_m+\frac{\lambda}{e^{\mu}}\Big\{(FL_m)''-
\Big(\mu'-3\nu^{'}\Big)\frac{(FL_m)'}{2}+\frac{2}{\tilde{r}}\Big(FL_m\Big)'\Big\}\Big]d\tilde{r}\Big)d\tilde{r}.\label{50}
\end{eqnarray}
Actually, the Eq.(\ref{50}) discussed the effects of local pressure anisotropy and
inhomogeneous energy density to the Tolman mass for the interior self-gravitating source in background of non-minimally
coupling curvature to matter with $f(R)$ gravity formalism.
\section{The Vanishing Condition of Complexity Factor}
In this evaluation numeral quantities are involved to produce complexity in a self-gravitating
system, such that anisotropy of the pressure, inhomogeneity of energy density, heat conduction and
viscosity parameters. In the absence of the above specific parameters we might introduce a well definite
general system that possess an energy density homogeneity and as well as pressure isotropy, then system
will have minor complexity. In our case, the system having pressure anisotropy and inhomogeneity of energy
density that provides the fundamental cause of the complexity in a system. These factors are involved in $Y_{TF}$
Eq.(\ref{48}) to the Tolman mass, which are responsible to generate the complexity factor in a system. Moreover,
to confer here in the attention of vanishing of complexity factor. We formulate the basic set of three ordinary
differential equations, which are formed by static spherically symmetric anisotropic self-gravitating geometry
accompanied with five unknown variables $(\nu,\mu,\rho,P_{r},P_{\bot})$ in context of non-minimal coupled curvature
matter with $f(R)$ theory of gravity. Therefore, we apply the condition $Y_{TF}=0$ and still one more condition is
required in this format to make the system is simple. Accordingly, by considering Eq.(\ref{48}) with the condition
of $Y_{TF}=0$, then the vanishing of complexity factor condition takes the following arrangement
\begin{eqnarray}\nonumber
&&\Pi=\frac{1}{2r^3(\frac{1+\kappa}{2})}\int^{r}_{0}\tilde{r}^3\Big(\kappa{'}\rho+\kappa \rho{'}\Big)d\tilde{r}-\frac{1}{2(\frac{1+\kappa}{2})}\int^{r}_{0}\frac{\kappa\tilde{r}^2}{(1+\lambda f(R))}\Big[-\lambda R F L_m\\\nonumber&&+2\lambda e^{-\mu}\Big\{(F L_m)''+ \Big(\frac{2}{\tilde{r}}-\frac{\mu^{'}}{2}\Big)(F L_m)'\Big\}\Big]d\tilde{r}+\frac{\kappa}{2(1+\lambda f(R))(\frac{1+\kappa}{2})}\Big\{-\lambda R F L_m\\&&+\frac{6\lambda (F L_m)'}{re^{\mu}}\Big\}.\label{51}
\end{eqnarray}
It is worth mentioning that the above expression reduces to Eq.(58) of Herrera \cite{9}, when $\lambda=0$.

Furthermore, in this pattern we present the example for the interior anisotropic source designed by spherically
symmetric structure with inconstant energy density that proposed by Gokhoo and Mehra \cite{52}. Actually, the
concept is used in the physical significance of the compact objects. In our account to use this concept for the relativistic
dense source, which explain the physical properties for the vanishing of complexity factor. The preliminary idea for getting
this object is to assume the metric function $\mu$ that has the following form
\begin{equation}\label{52}
e^{-\mu}=1-\alpha r^2+\frac{3r^{4}\Omega\alpha}{5r^{2}_{\Sigma}},
\end{equation}
where $\alpha=\frac{\rho_{0}}{3}$ and $\Omega$ treat as a constant in the interval $(0,1)$.

From Eqs.(\ref{10}) and (\ref{17}), which reads that
\begin{eqnarray}\nonumber
&&\kappa\rho=\rho_0\Big(1-\frac{\Omega r^2}{r^{2}_{\Sigma}}\Big)-\frac{\kappa\Big(1-\alpha r^2+\frac{3r^{4}\Omega\alpha}{5r^{2}_{\Sigma}}\Big)}{(1+\lambda f(R))}
\Big\{\frac{-\lambda R F L_m}{(1-\alpha r^2+\frac{3r^{4}\Omega\alpha}{5r^{2}_{\Sigma}})}\\&&+2\lambda(FL_m)''
+2\lambda\Big(\frac{2}{r}-\frac{(\alpha r-\frac{6 \Omega \alpha r^3}{5 r^{2}_{\Sigma}})}{(1-\alpha r^2+\frac{3r^{4}\Omega\alpha}{5r^{2}_{\Sigma}})}\Big)\Big(FL_m\Big)'\Big\},\label{53}
\end{eqnarray}
\begin{eqnarray}
&&m(r)=\frac{\rho_{0}r^3}{6}\Big(1-\frac{3\Omega r^2}{5r^{2}_{\Sigma}}\Big),\label{54}
\end{eqnarray}
we use Eqs.(\ref{11}) and (\ref{12}), it takes the form
\begin{eqnarray}\nonumber
&&\kappa\Pi(r)+\frac{1}{r^2}=\frac{1}{e^{\mu}}\Big[-\frac{\nu^{''}}{2}-\left(\frac{\nu^{'}}{2}\right)^2
+\frac{\nu^{'}}{2r}+\frac{1}{r^2}+\frac{\mu'}{2}\Big(\frac{1}{r}+\frac{\nu^{'}}{2}\Big)\Big]
+\frac{2\lambda\kappa}{e^{\mu}(1+\lambda f(R))}\Big[-(FL_m)''\\&&+\Big(\frac{\mu'}{2}+\frac{1}{r}\Big)\Big(FL_m\Big)'\Big],\label{55}
\end{eqnarray}
by considering the variables
\begin{equation}\label{56}
e^{\nu(r)}=e^{\int(2z(r)-\frac{2}{r})dr},
\end{equation}
and
\begin{equation}\label{57}
\frac{1}{e^{\mu}}=y(r).
\end{equation}
From the information of Eqs.(\ref{56}) and (\ref{57}) inserting in Eq.(\ref{55}), one can obtain
\begin{eqnarray}\nonumber
&&\Big[1+\frac{2\lambda\kappa\Big(FL_m\Big)'}{z\Big(1+\lambda f(R)\Big)}\Big]y'+y\Big[\frac{2z'}{z}+2z-\frac{6}{r}+\frac{4}{r^2z}+\frac{4\kappa\lambda}{z(1+\lambda f(R))}\Big(\Big(FL_m\Big)''-\frac{1}{r}\Big(FL_m\Big)'\Big)\Big]\\&&=-\frac{2}{z}\Big[\kappa\Pi(r)+\frac{1}{r^2}\Big].\label{58}
\end{eqnarray}
Consequently, we obtain the following line element in term of two functions $z$ and $\Pi$ which leads to \cite{9, 53},
these results are arranged in context of modified $f(R)$ gravity with nonminimally coupled to curvature matter.
\begin{eqnarray}\nonumber
&&ds^2=-e^{\int(2z(r)-\frac{2}{r})dr}dt^2\\\nonumber&&+\frac{z^2(r)e^{\int\Big[2z(r)+\frac{4}{r^2z(r)}+\frac{4\lambda\kappa}{z(r)(1+\lambda f(R))}\Big((FL_m)''-\frac{1}{r}(FL_m)'\Big)\Big]dr}}{r^6\Big[-2\int\frac{z(r)
\Big(\kappa\Pi(r)r^2+1\Big) e^{\int\Big[2z(r)+\frac{4}{r^2z(r)}+\frac{4\lambda\kappa}{z(r)(1+\lambda f(R))}\Big((FL_m)''-\frac{1}{r}(FL_m)'\Big)\Big]dr}}{r^8}dr+C\Big]}dr^2\\&&
+r^2d\theta^2+r^2sin^2\theta d\phi^2.\label{59}
\end{eqnarray}
Here $C$ is an integration constant.

Moreover, we the matter variables have following form
\begin{eqnarray}
\frac{\kappa P_{r}}{2}=\frac{z(r-2m)+\frac{m}{r}-1}{r^2}+\frac{\kappa(1-\frac{2m}{r})}{(1+\lambda f(R))}\Big[\frac{-\lambda R F L_m}{(1-\frac{2m}{r})}+2\lambda(z+\frac{1}{r})(FL_m)'\Big],\label{60}
\end{eqnarray}
\begin{eqnarray}
\frac{\kappa\rho}{2}=\frac{m'}{r^2}-\frac{\kappa(1-\frac{2m}{r})}{2(1+\lambda f(R))}\Big[\frac{-\lambda R F L_m}{(1-\frac{2m}{r})}+2\lambda \Big((FL_m)''+\Big(\frac{2}{r}-\frac{(\frac{m'}{r}-\frac{m}{r^2})}{(1-\frac{2m}{r})}\Big)(FL_m)'\Big)\Big],\label{61}
\end{eqnarray}
and
\begin{eqnarray}\nonumber
&&\kappa P_{\bot}=\Big(1-\frac{2m}{r}\Big)\Big(z'+z^2-\frac{z}{r}+\frac{1}{r^2}\Big)+z\Big(\frac{m}{r^2}-\frac{m'}{r}\Big)
+\frac{\kappa}{(1+\lambda f(R))}\Big[-\lambda R F L_m\\&&+2\lambda\Big(\Big(1-\frac{2m}{r}\Big)(FL_m)''+\Big(\Big(1-\frac{2m}{r}\Big)z+\Big(\frac{m}{r^2}-\frac{m'}{r}\Big)\Big)(FL_m)'\Big)\Big].\label{62}
\end{eqnarray}
Subsequently, to get most prominent invariant results in terms of radial pressure, energy density and tangential pressure that
follows to Eqs.(\ref{60})-(\ref{62}) in context of vanishing of complexity factor for the self-gravitating relativistic dense
structure.
\section{Conclusions}
In this symmetry, we signify the concept that was initially defined by Herrera \cite{9} for spherically symmetric
anisotropic distribution in framework of vanishing complexity factor. Our struggle is convenient to reformulate this phenomena
for the relativistic dense interior objects formed by static spherically symmetric anisotropic self-gravitating geometry in context
of non-minimal curvature coupling with $f(R)$ gravity theory for the vanishing of complexity factor. The term complexity is entirely
based on the structure scalars that comes from the orthogonal splitting of the Riemann-Cristofell tensor. The motive of this
theory is to expose the effects on the complexity factor either this complexity factor increasing or decreasing in a system.

In present work we examined such notable issues that introduced in our system. We have investigated the mass function
into two formats such as general mass \cite{47} and Tolman mass \cite {48}. These two functions have significant effects on the system and given in Eq.(\ref{30}) and Eq.(\ref{38}) respectively.
In the first case the distribution of the matter has homogeneous energy density and the variation induced in
the energy density inhomogeneity. The second case is about the pressure anisotropy and density energy inhomogeneity that
are calculated in term of Tolman mass.

Next, we have defined the scalar functions by the orthogonal splitting of the Riemann-Christofell
tensor and find the complexity factor in a system. The Eq.(\ref{48}) related to complexity factor
$Y_{TF}$, which includes the basic factors of pressure anisotropy and inhomogeneity energy density
that provides the physical cause of complexity factor in a system. We have concluded the complexity
factor vanishes with the assumption $Y_{TF}=0$ that are given in Eq.(\ref{51}). In other words the
system will have zero complexity when the effects of pressure anisotropy and density energy inhomogeneity
cancel to each other.

It is worth to mention that the insertion of the pressure anisotropy on $Y_{TF}$ is local whereas the insertion
of inhomogeneity energy density is not. We have also recognized in context of non-minimal coupled curvature matter with
metric $f(R)$ gravity theory agrees with pressure isotropy and homogeneous energy density to decrease the structure scalar $Y_{TF}$.
Furthermore, we analyzed the relativistic interior self-gravitating dense structure in the existence of specific energy density
with the vanishing complexity factor and the results are mentioned in Eqs.(\ref{60})-(\ref{62}).

In this interest, we would like to discuss that our results for $\lambda=0$ of Eqs.\Big((\ref{13})-(\ref{15}), (\ref{28})-(\ref{30}),
(\ref{34}), (\ref{38}), (\ref{48}), (\ref{50}), (\ref{51}), (\ref{54}), (\ref{59})-(\ref{62})\Big) approaches to
Eqs.\Big(9-11, 28-30, 33, 35, 54, 56, 58, 61, 66-69\Big) of Herrera Ref.\cite{9}.

The present article is referred to non-minimal coupled curvature matter with $f(R)$ metric theory in the background
of vanishing complexity factor,\textbf{ in future it would be interesting to reproduce this work with other phenomenal theories such as Rastall theory,
$f(T)$, $f(G)$, and $f(R,T)$ gravity theories}.
\section*{Acknowledgment}
One of us G.A appreciates the financial support from HEC, Islamabad, Pakistan under NRPU project with grant number 20-4059/NRPU/R \& D/HEC/14/1217. Also, we appreciate the constructive comments of anonymous referee that help a lot to improve the manuscript.
 
\end{document}